\documentstyle[revtex]{aps}
\begin{document}\preprint{OUPD9314}
\begin{title}
Local $^4$He--p potentials from RGM phase shifts
\end{title}
\author{S.~G. Cooper and R.~S. Mackintosh\cite{dagger}}
\begin{instit}
Physics Department, The Open University, Milton Keynes,
United Kingdom MK7 6AA
\end{instit}
\author{A. Cs\'ot\'o and R.~G. Lovas }
\begin{instit}
Institute of Nuclear Research of the Hungarian Academy of Sciences,\\
Debrecen, P. O. Box 51, H--4001, Hungary
\end{instit}
\receipt{30 August 1993}
\begin{abstract}
Phase shifts for $\alpha$ + nucleon scattering generated by a
multichannel RGM model of the five-nucleon system are subjected to
iterative-perturbative ``mixed-case" inversion for energies below the
reaction threshold. The resulting phase-equivalent potentials are
compared with local potentials calculated from the inversion of
empirical phase shifts. A strong similarity is revealed between the
two sets of potentials, most notably in the description by a parity-
and energy-dependent local potential. In particular, the RGM-derived
and empirical potentials share a distinctive form of parity
dependence in which the odd-parity component is of greater radial
extent. Comparison with potentials representing single-channel RGM
phase shifts exhibits the importance of the coupled channels in terms
of local potentials. The relative wave functions derived from RGM
are very different to those for the phase-equivalent local potentials.
\end{abstract}

\section{Introduction}
\label{sec:intro}
Although nuclei are composite objects, it is still meaningful to
describe their interaction with a nucleon or other nucleus by a local
two-body potential. This is the idea underlying the optical model,
which continues to play a role in our understanding of nuclear
scattering. In an attempt to incorporate the full information content
of the empirical data and at the same time avoid the serious
ambiguity problems besetting conventional elastic scattering
phenomenology, a two-step approach has been introduced. The data is
first fitted by phase shifts; these empirical phase shifts are then
``inverted" to yield potentials which closely reproduce
them~\cite{CMNP517,CMPRC40,CMPRC43,CMPRC45,MCMNP552}. The cited
calculations all use the iterative-perturbative (IP)
method~\cite{CMInverse,CMNP511,CMNP513,CMZP} outlined in
Sec.~\ref{sec:inversion} below. The resulting potentials are
generally better determined than those from conventional
optical-model searches, behave reasonably as functions of energy and
angular momentum, and also have characteristic properties which elude
conventional phenomenology.

Well determined local potentials motivate us to establish their
meaning and study their relation to a microscopic description
involving nucleonic degrees of freedom. In particular, which
characteristic properties of the empirical local potentials are
predicted by the microscopic description? The system $^4$He+p
considered here is very suitable for a study of this kind for three
reasons: the empirical potentials have been deduced from good-quality
data, they behave in a non-trivial manner, and realistic microscopic
calculations are feasible.

In this paper we apply inversion to deduce local one-body potentials
from $^4$He+p phase shifts calculated from a microscopic scattering
model. We then compare these with local potentials determined by the
two-step method and known~\cite{CMPRC43} to give a good
representation of experimental data. By examining the wave functions
from the two models, we shall gain some measure of the effect of
representing a well-founded non-local model by local potentials.

The microscopic model we use is of the resonating-group (RGM) type,
with two partitions, \{$\alpha$+p,$^3$He+d\} or \{$\alpha$+n,t+d\}.
The local potential representing the RGM phase shifts is obtained by
means of ``mixed-case" inversion using the IP
algorithm~\cite{CMInverse}. In mixed-case
inversion~\cite{CMNP517,CMPRC43}, potentials are constructed which
simultaneously reproduce phase shifts for a set of angular momenta
and a range of energies. The inversion is appropriate to
spin-${1\over 2}$ particles, and  spin-orbit potentials are
presented. The present calculations are concerned only with the
energy range below the $^3$He+d (t+d) threshold, so all potentials
are real.

The most interesting features of the empirical (two-step) potential
are a strong parity-dependent shape and a spin-orbit term that is of
shorter range than the central term. We establish whether a potential
found by inverting RGM phase shifts has these same properties and
thereby judge whether these properties arise from the basic
properties of the RGM theory. It has been known for a long time that
Pauli exchange built into the RGM leads to parity dependence, but up
to now there has been no way of knowing how this should be
represented in a potential model; generally it has been assumed that
the shape of the potential is not parity dependent. Other properties
of the RGM procedure can also be studied. For example, by omitting
the coupling to the deuteron channel, we get a clear measure of the
influence of such coupling on a potential model for $^4$He+p
scattering.

Since inversion leads to a local potential having exactly the RGM
phase shifts, the two models are ``phase equivalent'' and it is
desirable to compare the RGM and local potential wave functions in
the interaction region. To do this, we must eliminate the ambiguity
caused by the mixing of the relative and intrinsic motions in the RGM
wave function due to antisymmetrization. We do this by introducing a
most reasonable definition for the relative wave function. In this
way the RGM formalism will provide us with trivially equivalent local
potentials (TELP's) and  permit the desired comparison of wave
functions. An examination of the resulting so-called Perey factors
provides a measure of the non-locality effects.

In Sec.~\ref{sec:formalism} we shall summarize the microscopic model
(\ref{subsec:model}), outline the foundations of its relationship to
a macroscopically meaningful relative motion
(Sec.~\ref{subsec:macrmod}), and introduce the TELP's
(Sec.~\ref{subsec:TELP}). In Sec.~\ref{sec:inversion} we shall
recapitulate the essentials of the inversion procedure applied. We
present the results in Sec.~\ref{sec:results}. In this we shall start
with the RGM results (Sec.~\ref{subsec:phaseshifts}), then we present
the inversion potentials deduced from the RGM phase shifts
(Sec.~\ref{subsec:potentials}), and, finally, we shall show some
Perey factors and trivially equivalent potentials
(Sec.~\ref{subsec:Perey}). This will be followed by some general
discussion and conclusions in Sec.~\ref{discussion}.

\section{Microscopic formalism}
\label{sec:formalism}
\subsection{The microscopic model}
\label{subsec:model}
We use an approximation to an RGM based on a gen\-er\-ator-coordinate
method (GCM). In a coupled-channel (CC) version of the model of the
$^5$Li-like system the wave function for a particular total angular
momentum $j$ and parity $\pi$ is a combination of the $\alpha+p$ and
$h+d$ ($h=^3$He) partitions:
\begin{equation}
\Psi_j^{\pi}=\Psi^{(\alpha p)}_{[l{1\over 2}]j}
+\sum_{l'}\sum_{s'={1\over 2},{3\over 2}}\Psi^{(hd)}_{[l's']j},
\label{CCtrialfn}
\end{equation}
where $l=j\pm{1\over 2}$ with $(-1)^l=(-1)^{l'}=\pi$ and
\begin{equation}
\Psi^{(12)}_{[ls]j}
={\cal A}_{12}\left \{\left[[\Phi^{(1)}_{s_1}\Phi^{(2)}_{s_2}]_s
\varphi^{(12)}_{l}({\bf r}_{12})\right]_{j}\right\}.
\label{SCtrialfn}
\end{equation}
Here the functions $\Phi$ are antisymmetrized intrinsic wave
functions (or the spin--isospin function for the nucleon),
$[\ ]_{JM}$ denotes angular-momentum coupling, and
${\cal A}_{12}$ is an inter-fragment antisymmetrizer. We define
${\cal A}_{12}$ as
\begin{equation}
{\cal A}_{12}=N^{-1/2}\sum^N_{i\in 1,j\in 2}(-1)^{P_{ij}}P_{ij},
\label{antisymmetrizer}
\end{equation}
where $N=(A_1+A_2)!/A_1!A_2!$ is the number of permutations $P_{ij}$
that mix the particle labels in fragment 1 with those in 2.

We describe the internal motions of $\alpha$ and $h$ by single
translation-invariant 0s oscillator shell-model configurations and
that of $d$ by a combination of three such functions of different
size parameters. For instance,
\begin{equation}
\Phi^{\alpha}_0=\frac{1}{\sqrt{4!}}\frac
{\psi^{(\beta)}_{000}({\bf r}_1)...
\psi^{(\beta)}_{000}({\bf r}_4){\cal X}}
{\psi^{(4\beta)}_{000}(\textstyle{\frac14}
({\bf r}_1+...+{\bf r}_4))},
\label{alphawf}
\end{equation}
where $\psi^{(\beta)}_{000}({\bf r})
=(\beta/\pi)^{3/4}\exp(-\frac12\beta r^2)$
and $\cal X$ is the spin-isospin factor. The functions that describe
the relative motions are expanded, as in the conventional GCM, in
terms of angular-momentum projected shifted Gaussians,
\FL
\begin{equation}
\varphi^{(12)}_{lm}({\bf r})=\sum _k C_k
\hbox{${\displaystyle \int}\kern -0.3em d$}\hat{\bf R}_k
Y_{lm}(\hat {\bf R}_k)\left({\gamma\over\pi}\right)^{3/4}
e^{-{1\over 2}\gamma({\bf r}-{\bf R}_k)^2},
\label{expansion}
\end{equation}
matched with S-matrix asymptotics~\cite{Kamimura}. The width $\gamma$
and the radial centres $\{R_k\}$ of the Gaussians involved in the
expansion can be fixed easily so as to make the basis set dense
enough. The expansion coefficients are determined by a variational
approximation to the five-particle problem, and the S-matrix is
calculated with the Kohn--Kato formula~\cite{Kamimura}.

The nucleon--nucleon interaction in the Hamiltonian was chosen
to have the form
\FL
\begin{eqnarray}
V(i,j)&=&({\cal P}_tV_t{\rm e} ^{-r^2/a_t^2}
+{\cal P}_sV_s{\rm e} ^{-r^2/a_s^2})
{\textstyle {1\over 2}}[u+(1-u)P^r_{ij}]
\nonumber\\
&&+(W_T+M_TP_{ij}^r)r^2\nonumber\\
&\times&\sum _{k=1,2}
V_{{\rm T}_k}{\rm e} ^{-r^2/a_{{\rm T}_k}^2}
[3(\mbox{\boldmath $\sigma $}_i\cdot{\bf r})
(\mbox{\boldmath $\sigma $}_j\cdot{\bf r})/r^2
-\mbox{\boldmath $\sigma $}_i\cdot\mbox{\boldmath $\sigma $}_j]
\nonumber\\
&&+V_{\rm s.o.}{\rm e} ^{-r^2/a_{\rm s.o.}^2}\hbar ^{-1}
{\bf l}\cdot(\mbox{\boldmath $\sigma $}_i
+\mbox{\boldmath $\sigma $}_j),
\label{pot}
\end{eqnarray}
where ${\cal P}_t$ and ${\cal P}_s$ are spin-triplet and -singlet
projectors, respectively, $P^r_{ij}$ is the space-exchange operator,
${\bf r}={\bf r}_j-{\bf r}_i$, {\boldmath $\sigma $}$_i$ are the
Pauli vectors of the nucleonic spin, and
\begin{equation}
{\bf l}=-{\textstyle {1\over 2}} i\hbar {\bf r}
\times (\nabla _j-\nabla _i)
\end{equation}
is the orbital momentum of the relative motion of the two nucleons.
The calculation of the matrix elements involved has been performed
analytically.

Just as in our earlier work~\cite{CsLK}, we took the parameters from
Bl\"uge and Langanke~\cite{BLPR,BLFB}.
This interaction combines the central term of the force constructed
by Chwieroth {\it et al.}~\cite{CTT} with
a spin-orbit force of Reich\-stein and Tang~\cite{RT},
($V_{\rm s.o.}=-224.8$ MeV and
$a_{\rm s.o.}=0.707$ fm) and a slightly modified
version of the tensor interaction of Heiss and
Hackenbroich~\cite{HH}. We set the free space-exchange parameter $u$
to $0.835$~\cite{BLPR,BLFB}. The proton--proton Coulomb potential was
approximated by a combination of 15 Gaussians~\cite{PLNGyV}.
The oscillator parameters that we adopted are those given by
Chwieroth {\it et al.}~\cite{CTT} and these generate clusters of the
correct size.

For each quantum number $j^{\pi}$ we included all combinations of
$l's'$ that are allowed. Unlike in Ref.~\cite{CsLK}, the deuteron
distortion (or excitation) is not now allowed for because the force
parameters adopted turn out to reproduce the $\alpha+p$ empirical
phase shifts at low energies better in this way. To express
the fact that the excitations of the clusters are suppressed, whether
their ground states are taken as single or mixed configurations, we
refer to our models as {\em frozen\/}~\cite{LKL}.
It is clear that the frozen model seems to work better because the
parameters of the central term of the force have been fitted to the
$\alpha+p$ phase shifts in the framework of such a model~\cite{CTT}.

We repeated the calculations in a single-channel (SC) $\alpha+p$
model, which omits the second term of Eq.~(\ref{CCtrialfn}) from the
CC model. We also performed similar calculations for the mirror
system $\alpha+n$.

\subsection{Derivation of a macroscopic model}
\label{subsec:macrmod}
The local $\alpha+p$ potential to be found by inversion constitutes a
macroscopic model involving a structureless $\alpha$ particle and a
proton interacting via a hermitian two-body interaction. It is only
within the SC framework that it is straightforward to establish a
correspondence to the microscopic approach. The structureless
$\alpha$ in the macroscopic approach implies a non-excitable $\alpha$
particle in the corresponding microscopic approach. In such an SC
microscopic approach the five-nucleon wave function can be expanded
as
\begin{equation}
\Psi=\hbox{${\displaystyle \int}\kern -0.3em d$}{\bf r}
\varphi ({\bf r})\Psi_{\mbox{\scriptsize{\bf r}}},
\label{Psi}
\end{equation}
where, for each value of {\bf r}, the function
$\Psi_{\mbox{\scriptsize{\bf r}}}$  is the element of
a basis, labelled by the continuous index {\bf r},
\FL
\begin{equation}
\Psi_{\mbox{\scriptsize{\bf r}}}
(\xi _{\alpha },\xi _p,{\bf r}_{\alpha p})
={\cal A}_{\alpha p}\{\Phi^{(\alpha)}(\xi _{\alpha })
\Phi^{(p)}(\xi _p)
\delta ({\bf r}-{\bf r}_{\alpha p})\},
\label{Psir}
\end{equation}
and $\varphi ({\bf r})$ is the corresponding expansion coefficient.
(From now on, for simplicity, angular-momentum coupling is
suppressed.) Since Eqs.~(\ref{Psi},\ref{Psir}) formulate the standard
cluster-model ansatz, we can follow the cluster-model formalism in
the ensuing exposition~\cite{Saito}. Projected onto this subspace,
the five-particle Schr\"odinger equation, $H\Psi=\varepsilon\Psi $,
reduces to
\begin{equation}
{\cal H}\varphi ({\bf r})=\varepsilon{\cal N}\varphi ({\bf r}),
\label{Hphi=ENphi}
\end{equation}
where ${\cal H}$ and ${\cal N}$ are integral operators whose kernels
are
\begin{mathletters}
\begin{eqnarray}
H({\bf r},{\bf r'})&=&\langle \Psi_{\mbox{\scriptsize{\bf r}}}\vert H
\vert \Psi_{\mbox{\scriptsize{\bf r}}'}\rangle,
\label{Hkernel}
\\
N({\bf r},{\bf r'})&=&
\langle \Psi_{\mbox{\scriptsize{\bf r}}}\vert
\Psi_{\mbox{\scriptsize{\bf r}}'}\rangle .
\label{Nkernel}
\end{eqnarray}
\end{mathletters}
[E.g., ${\cal N}f({\bf r})\equiv \int {\rm d }{\bf r'}\
N({\bf r},{\bf r'})f({\bf r'})$.] In Eq.~(\ref{Hphi=ENphi}) we had to
have recourse to operators acting on the parameter
coordinate {\bf r}. In this way we managed to introduce relative
variables without violating the Pauli principle. The effect of the
Pauli principle is incorporated in the definitions of the kernels.

The appearance of ${\cal N}$ on the right hand side requires
Eq.~(\ref{Hphi=ENphi}) to be rewritten in some way before it can be
identified with a two-particle Schr\"odinger equation. One
possibility would be to write
\begin{equation}
[{\cal H}+({\bf 1}-{\cal N})\varepsilon]\varphi ({\bf r})
=\varepsilon\varphi({\bf r}), \label{Schrforphi}
\end{equation}
with the operator in the square brackets considered a redefined
Hamiltonian. Alternatively, one can multiply Eq. (\ref{Hphi=ENphi})
from the left by ${\cal N}^{-1}$ and regard ${\cal N}^{-1}{\cal H}$
as the Hamiltonian. For this inverse operation to be permissible, one
should exclude the subspace $\{\varphi_i^{(0)}\}$ of $\{\varphi \}$
for which ${\cal N}\varphi _i^{(0)}=0$. This is not an actual
restriction because ${\cal N}\varphi _i^{(0)}=0$ implies
${\cal A}_{\alpha p}\{\Phi^{(\alpha)}\Phi^{(p)}
\varphi _i^{(0)}({\bf r}_{\alpha p})\}=0$, and thus $\varphi $ and
$\varphi +\sum _ic_i\varphi _i^{(0)}$ generate the same five-particle
state. The null effect of $\varphi^{(i)}_0$ in $\Psi$ reflects the
fact that $\varphi^{(i)}_0$ are Pauli-forbidden relative-motion
states. The exclusion of the subspace $\{\varphi _i^{(0)}\}$ implies
that ${\cal N}$, having no zero eigenvalue on the restricted
(Pauli-allowed) state space of relative motion, can be inverted. In
fact, in our problem there is just one state $\varphi^{(0)}$, a 0s
oscillator state.

The Hamiltonian ${\cal H}+({\bf 1}-{\cal N})\varepsilon$ is, however,
strongly energy dependent, and the other Hamiltonian,
${\cal N}^{-1}{\cal H}$, is non-hermitian. Correspondingly, the
normalization of the function $\varphi$ is related to that of the
wave functions $\Psi$ as
$(\varphi\vert{\cal N}\vert\varphi ')=\langle\Psi\vert\Psi '\rangle$,
which shows a departure from the normalization of a wave function.
(The round brackets denote matrix elements involving integrations
over the parameter coordinates.) Thus none of these is the proper
reduction.

One can obtain a two-particle Schr\"odinger equation with a hermitian
as well as energy-independent Hamiltonian by multiplication of
Eq.~(\ref{Hphi=ENphi}) from the left by ${\cal N}^{-1/2}$ and
redefining the wave function and the Hamiltonian as
\begin{mathletters}
\begin{eqnarray}
\chi ({\bf r})&=&{\cal N}^{1/2}\varphi ({\bf r}),
\label{relwf}
\\
h&=&{\cal N}^{-1/2}{\cal H}{\cal N}^{-1/2},
\label{relHam}
\end{eqnarray}
\end{mathletters}
with the result
\begin{equation}
h\chi ({\bf r})=\varepsilon\chi ({\bf r}).
\label{relSchr}
\end{equation}
This transformation implies for the relative
wave function the same normalization as is valid for $\Psi $:
\begin{equation}
\langle\Psi\vert\Psi'\rangle
=(\varphi\vert{\cal N}\vert\varphi')=(\chi\vert\chi').
\label{norm}
\end{equation}
This is a necessary condition for a function to be interpreted as a
wave function, assigning it probability meaning. The
wave-function-like property of $\chi$ is emphasized by the
fact~\cite{HoriuchiPerey} that, from among the Hamiltonians $\cal H$,
$h$, and ${\cal H}{\cal N}^{-1}$, whose eigensolutions are $\varphi$,
$\chi$, and $\eta={\cal N}\varphi$, respectively, that belonging to
$\chi$ can be best approximated by a Hamiltonian containing a local
potential.

Before fully establishing the correspondence between the microscopic
and macroscopic models, we make more explicit the relation of the
Hamiltonian $h$ of the two-particle model to that of the five-nucleon
system. To this end, let us write the microscopic Hamiltonian, in
self-explanatory notation, in the asymmetric form
\begin{equation}
H=H_{\alpha }(\xi _{\alpha })+T_{\alpha p}({\bf r}_{\alpha p})
+V_{\alpha p}(\xi _{\alpha },{\bf r}_{\alpha p}).
\label{Hasymm}
\end{equation}
Substituting (\ref{Psir}) and (\ref{Hasymm}) into Eq.~(\ref{Hkernel})
and using ${\cal A}_{\alpha p}^2=\sqrt{5}{\cal A}_{\alpha p}$, we
arrive at
\begin{equation}
H({\bf r},{\bf r'})=[E_{\alpha }+T_{\alpha p}({\bf r})]
N({\bf r},{\bf r'})+V_{\alpha p}({\bf r},{\bf r'}),
\label{Hkernasymm}
\end{equation}
where $E_{\alpha }$ is the ground-state energy of the $\alpha$
particle and
\begin{mathletters}
\FL
\begin{eqnarray}
V_{\alpha p}({\bf r},{\bf r'})
&=&\sqrt{5}\langle \Phi^{(\alpha)}\Phi^{(p)}
\delta ({\bf r}-{\bf r}_{\alpha p})\vert
V_{\alpha p}\vert \Psi_{\mbox{\scriptsize{\bf r}}'}\rangle
\label{nonlocpota}\\
&=&\langle {\cal A}_{\alpha p}\{\Phi^{(\alpha)}\Phi^{(p)}
\delta ({\bf r}-{\bf r}_{\alpha p})
V_{\alpha p}\}\vert \Psi_{\mbox{\scriptsize{\bf r}}'}\rangle .
\label{nonlocpotb}
\end{eqnarray}
\end{mathletters}
Denoting the integral operators belonging to the kernels by the
corresponding script letters, we can express $h$ as
\begin{mathletters}
\begin{eqnarray}
h&=&E_{\alpha }+{\cal N}^{-1/2}T_{\alpha p}{\cal N}^{1/2}
+{\cal N}^{-1/2}{\cal V}_{\alpha p}{\cal N}^{-1/2}
\label{hasymma}\\
&\equiv&E_{\alpha}+T_{\alpha p}+v,
\label{hasymmb}
\end{eqnarray}
\end{mathletters}
where $v$ is the potential that is the best candidate to correspond
to the macroscopic potential. Since $T_{\alpha p}$ does not commute
with ${\cal N}$, the local potential $v$ contains a contribution from
the kinetic energy as well. This contribution is, however, expected
to be small~\cite{Timm}, so that
\begin{equation}
v\approx{\cal N}^{-1/2}{\cal V}_{\alpha p}{\cal N}^{-1/2}.
\label{vpot}
\end{equation}
With the potential $v$ in Eqs.~(\ref{hasymmb},\ref{vpot}), the
Schr\"odinger equation for the relative motion, Eq.~(\ref{relSchr}),
reads
\begin{equation}
(T_{\alpha p}+v)\chi=E\chi \ \ \ \ \ \ \ \ \ \ \ \ \ \ \ \
(E=\varepsilon-E_{\alpha}).
\label{relSchrwithpot}
\end{equation}

The transformation (\ref{relwf},\ref{relHam}) is the only one that
produces a hermitian effective Hamiltonian, apart from unitary
transformations
${\cal U}$ applied to $\chi$ and $h$:
\begin{equation}
\tilde\chi({\bf r})={\cal U}\chi({\bf r}),\ \ \ \ \ \ \ \ \ \
\tilde h={\cal U}h{\cal U}^{-1}.
\label{unitrans}
\end{equation}
Because of the unitarity of ${\cal U}$, $\tilde h$ is also hermitian,
$\tilde\chi$ is properly normalized,
$\langle\chi\vert{\cal U}^{\dag}{\cal U}\vert\chi\rangle
=\langle\chi\vert\chi\rangle$, and, if (\ref{relSchr}) holds, so does
\begin{equation}
\tilde h\tilde\chi({\bf r})=\varepsilon\tilde\chi({\bf r}).
\label{altSchr}
\end{equation}
If, in addition, ${\cal U}$ is of finite range, then
$\tilde\chi({\bf r})=\chi({\bf r})$ is implied for the asymptotic
region, whence it follows that the potential in $\tilde h$ is a phase
equivalent of that involved in $h$. The function $\chi $ is
distinguished from the infinite number of functions $\tilde\chi$ not
only by aesthetics. An example shows~\cite{Schmid} that in matrix
elements the many-particle wave function is represented faithfully
just by $\chi $.

But before declaring that Eq.~(\ref{relSchrwithpot}) is the
Schr\"{o}dinger equation that
corresponds to the macroscopic local potential problem, we should
examine whether it is the only one. To this end, we should see
whether there is a finite-range unitary transformation $\cal U$
transforming $v$ into another ``almost local" potential. In fact,
what one can prove precisely is that there exists no non-trivial
transformation of this type between two {\em local} potentials. Let
$V({\bf r})$ be a local potential and $u({\bf r},{\bf r}')$ and
$u^{-1}({\bf r},{\bf r}')$, respectively, be the kernels of a unitary
transformation $\cal U$ and of its inverse, and let us assume that
${\cal U}V{\cal U}^{-1}$ produces a local potential $\tilde V$:
\begin{equation}
\hbox{${\displaystyle \int}\kern -0.3em{\rm d}$}{\bf r}'
u({\bf r},{\bf r}')V({\bf r}')u^{-1}({\bf r}',{\bf r}'')
=\tilde V({\bf r})\delta({\bf r}-{\bf r}'').
\label{transformation}
\end{equation}
By multiplying both sides by $u({\bf r}'',{\bf r}''')$ and
integrating over ${\bf r}''$, we obtain
\begin{equation}
u({\bf r},{\bf r}''')V({\bf r}''')
=\tilde V({\bf r})u({\bf r},{\bf r}'''),
\label{nonexistence}
\end{equation}
which can hold either if
$u({\bf r},{\bf r}')=\delta({\bf r}-{\bf r}')$ (i.e.
${\cal U}=\openone$) and then, obviously,
$\tilde V({\bf r})=V({\bf r})$, or else if
$V({\bf r}''')=\tilde V({\bf r})=$const. Thus, if $v$ were exactly
local, it would be unique. Although it is not perfectly local, it is
likely that is can be approximated by hermitian local
phase-equivalents. On these grounds, we shall consider
Eq.~(\ref{relSchrwithpot}) the Schr\"odinger equation that
corresponds to the macroscopic model.

Derivations of macroscopic potentials from the RGM kernels invariably
yield deep potentials, which accommodate states whose quantum numbers
are forbidden by the Pauli principle~\cite{Friedrich,Horiuchi}. Since
it is these potentials that produce eigenfunctions comparable with
the function $\chi$ of the RGM~\cite{SchmidNP,LovasPal}, one cannot
but accept that these are the local equivalent potentials that are
correct off the energy shell as well. The Pauli-forbidden states are
then to be eliminated by a projector included in the Schr\"odinger
equation. Depending on the derivation, either the Pauli forbidden
states defined by the RGM are to be projected out~\cite{Friedrich},
which leads to the orthogonality-condition model~\cite{Saito}, or the
Pauli-forbidden states may be identified with the low-lying states
generated by the local potential itself~\cite{Buck,Horiuchi}, which
reduces to a purely local-potential model with discarded states. The
procedure of explicit Pauli projecting cannot be avoided, however, if
these potentials are applied to systems of more than two bodies.

We have now argued that there should exist local Hamiltonians whose
eigenfunctions approximate the RGM wave function $\chi$, but we did
not exclude the existence of local potentials that are phase
equivalent but do not produce the correct wave function in the
interaction region, i.e. are correct only on the energy shell. In
fact, given a local potential, there exists a method, the so-called
supersymmetric transformation, to construct such phase-equivalent
potentials \cite{Baye}. (This transformation is not invertible and
hence is not unitary.) The supersymmetric partner of a potential
produces a wave function with one less node. Since, however, a
supersymmetric partner of a well-behaved potential has an $r^{-2}$
singularity at the origin, it is easy to avoid constructing such
potentials in an inversion procedure. It should be noted that the
supersymmetric partner has been shown to be much inferior to the
deep potentials in reproducing the predictions of the RGM in
wave-function-sensitive properties~\cite{liu}.

We should mention here that
these points are disregarded in the two works that are closest in
their objectives to ours, Kanada {\em et
al.}~\cite{Kanada} and the recent work of Howell {\em et
al.}~\cite{Howell}. Thus the TELP's
produced by Kanada {\em et al.} are hermitian local equivalents of
non-hermitian or energy-dependent non-local potentials which produce
``wave functions" of no probability meaning. The philosophy of Howell
{\em et al.}, whose work is restricted to the $S$-wave only, is also
different. They replace a local phase equivalent of an RGM-based
non-local potential, which could as well be considered a local
approximant of $v$, by its supersymmetric partner, in order to
eliminate the Pauli-forbidden state, and hence obtain a potential
applicable in an ordinary Schr\"odinger equation containing no
provision for Pauli projection. This means, however, that their wave
function is deprived of an inner node,
making it qualitatively different from
the correct RGM wave function. This will render the
structure calculations, for which they recommend their potential
(e.g. in the description of $^6$He as $\alpha+n+n$), unphysical in
this respect.

\subsection{Trivially equivalent local potentials}
\label{subsec:TELP}
The microscopic potential $v$ is an integral operator of a
finite-range kernel, i.e. it is non-local. A TELP $U$ of a non-local
potential $v$ produces the same wave function $\chi$ in the equation
\begin{equation}
(T_{\alpha p}+U)\chi=E\chi
\label{locSchr}
\end{equation}
as $v$ in Eq.~(\ref{relSchrwithpot}). We shall construct two types of
TELP's: one $l$-independent and one $l$-dependent.

We come to the $l$-independent TELP by taking
\begin{equation}
U(\mbox{${\bf r}$})\equiv
V(\mbox{${\bf r}$})+iW(\mbox{${\bf r}$})
=[\chi({\bf r})]^{-1}v\chi({\bf r}).
\label{TELP}
\end{equation}
With this function substituted, Eq.~(\ref{locSchr}) boils down to
(\ref{relSchrwithpot}), which verifies that such a local potential
yields the same relative-motion wave function $\chi$ in the entire
space. It is clear that it must be complex and, through the energy
dependence and boundary condition implicit in $\chi$, it must depend
on the energy and on the direction of $\mbox{${\bf r}$}$. However, it
contains no $l$-dependence, and, apart from the pathological case of
all partial waves having nodes at the same position, it is not
singular.

It is easy to see that the definition (\ref{TELP}) is identical to
the definition of the $\psi$-potential introduced in Ref.~\cite{MIC}.
Indeed, with Eq.~(\ref{relSchrwithpot}), the function $U$ can be cast
into the form $\chi^{-1}(E-T_{\alpha p})\chi$, which implies
\begin{mathletters}
\begin{eqnarray}
V(\mbox{${\bf r}$})&=&E+{\hbar^2\over 2\mu}
{{\rm Re}[\chi^*(\mbox{${\bf r}$})\nabla^2\chi(\mbox{${\bf r}$})]
\over\vert\chi(\mbox{${\bf r}$})\vert^2},
\label{repsipot}\\
W(\mbox{${\bf r}$})&=&{\hbar^2\over 2\mu}
{{\rm Im}[\chi^*(\mbox{${\bf r}$})\nabla^2\chi(\mbox{${\bf r}$})]
\over\vert\chi(\mbox{${\bf r}$})\vert^2}
\equiv{\hbar\over 2}{\nabla\cdot\mbox{${\bf j}$}(\mbox{${\bf r}$})
\over\vert\chi(\mbox{${\bf r}$})\vert^2}
\label{impsipot},
\end{eqnarray}
\end{mathletters}
where $\mu$ is the reduced mass of the
fragments and
$\mbox{${\bf j}$}(\mbox{${\bf r}$})$ is the vector of the probability
current. Eqs.~(\ref{repsipot},\ref{impsipot}) are equivalent to those
defining the $\psi$-potential in Ref.~\cite{MIC}.

Note that if $\chi$ were a solution of a problem with a real local
potential, then, because of $W\equiv 0$, the current would be free of
divergence. However, the W constructed for a non-local problem does
not vanish, which shows that a non-local potential pumps the flux
from one location to another.

By analogy with Eq.~(\ref{TELP}), one can define a TELP for each
partial wave:
\begin{equation}
U_l(r)=[\chi_l(r)]^{-1}v_l\chi_l(r),
\label{lTELP}
\end{equation}
where the radial functions and kernels are
\begin{eqnarray}
\chi_l(r)&=&r\hbox{${\displaystyle \int}\kern -0.3em
{\rm d}$}\hat{\mbox{${\bf r}$}}Y_{l\lambda}^*(\hat{\mbox{${\bf r}$}})
\chi(\mbox{${\bf r}$}),
\label{chil}
\\
v_l(r,r')&=&rr'\hbox{${\displaystyle \int}\kern -0.3em
{\rm d}$}\hat{\mbox{${\bf r}$}} Y_{l\lambda}(\hat{\mbox{${\bf r}$}})
\hbox{${\displaystyle\int}\kern-0.3em{\rm d}$}\hat{\mbox{${\bf r'}$}}
Y_{l\lambda}^*(\hat{\mbox{${\bf r}$}'})
v(\mbox{${\bf r}$},\mbox{${\bf r'}$}).
\label{vl}
\end{eqnarray}
It is immediately seen that $U_l(r)$ is not only $l$-dependent, but
also energy dependent and may be singular at the nodes of
$\chi_l(r)$, but does not depend on a direction. Similarly,
carrying through the angular-momentum coupling in $\Psi$,
one can easily derive an $lj$-dependent radial function, $\chi_{lj}$.

\section{The inversion procedure}
\label{sec:inversion}

We obtain potentials, local equivalent to the microscopic phase
shifts, by applying the iterative-perturbative (IP) method for
inversion implemented in the code IMAGO~\cite{imago}. The extension
to mixed-case inversion, in which we fit phase shifts corresponding
to a limited set of partial waves over a limited energy range, was
introduced in Ref.~\cite{CMNP517}. In mixed-case inversion particular
subsets of partial waves may be included, for example to
produce either energy-independent $l$-dependent potentials or
energy-dependent $l$-independent potentials, within the constraints
imposed by the limits of the data available. Here we closely follow
the techniques, developed in  Ref.~\cite{CMPRC43}, in which we
introduced the concept of energy bites. The idea is to perform the
inversion upon phase shifts for a set of closely spaced energies in
order to obtain a potential  reproducing, for each partial wave, both
the mean phase shift across the  energy interval and the energy
dependence of the phase shift. This technique is essential to
stabilize  what is effectively fixed-energy inversion for low-energy
scattering where few phase shifts significantly differ from zero. In
Ref.~\cite{CMPRC43}, use of these techniques enabled the derivation
of energy- and parity-dependent potentials from empirical phase
shifts for low-energy $\alpha+p$ scattering.

As the IP method has been previously described, here we only note the
relevant points and definitions which we refer to in the following
section. The algorithm iteratively corrects a ``starting reference
potential" by adding a linear superposition of basis functions. Since
the calculations presented here are obtained with a very small basis,
we expect some sensitivity to the choice of the basis. This parameter
dependence is assessed by performing otherwise identical calculations
using either a Gaussian or a zeroth order Bessel function basis. The
quality of an inversion is quantified by a measure referred to as the
``phase-shift distance", $\sigma$. This is defined in terms of the
target phase shifts $\{\delta^{\rm T}(l,j,E)$\}, for which we seek
the underlying potential, and the phase shifts
$\{\delta^{\rm I}(l,j,E)$\} calculated from that potential found by
inversion, as follows:
\FL
\begin{equation}
\sigma^2 = \sum_{l,j,E}\left\vert\exp[2i\delta^{\rm T}(l,j,E)]
-\exp[2i\delta^{\rm I}(l,j,E)]\right\vert^2,
\end{equation}
where the sum is over all required $l,j$ values and over all values
of $E$ defined in the energy bite.

Even with the techniques of mixed-case inversion outlined above,
the fact that there are at most four contributing $l$-values
implies that one can use
only a relatively small inversion basis. This could lead to
ambiguities, which must be minimized by searching for the smoothest
possible potentials and, in the case of energy-dependent potentials,
for a regular energy dependence. Inversion can lead to potentials
with oscillatory features; these features may arise as a result of an
underlying $l$-dependence or non-locality, but may also simply
represent artifacts of the inversion procedure. We have reduced the
presence of such artifacts by a simple smoothing procedure which has
been described in Ref.~\cite{CMPRC43}. In essence, this involves
replacing long-range surface oscillations in the inversion potentials
with a more physical exponential tail. If this smoothing does not
lead to convergence in the inversion procedure, we infer that the
oscillations are necessary to fit the given selection of partial
waves. This is typical for $\delta(l,j)$ derived from $l$-dependent
or energy-dependent potentials.

The inversion calculations for the proton case include the usual
Coulomb potential, based on a uniform spherical charge, customary in
optical model calculations. This facilitates the subsequent use of
our nuclear potentials in optical model codes. It follows that the
nuclear potentials which we present will have a small error, near the
nuclear surface, equal to the difference between the conventional and
exact Coulomb potentials. In all calculations presented
here the conventional charge radius used was $1.3 \times 4^{1/3}$ fm.
All inversion calculations were performed with a matching radius of 8
fm.

\section{Results}
\label{sec:results}
\subsection{Microscopic phase shifts}
\label{subsec:phaseshifts}

The phase shifts obtained from the RGM CC and SC calculations are
presented in Fig.~\ref{fig1} and Fig.~\ref{fig2} as functions of
energy. (All energy values are given in the laboratory frame.) The
empirical phase shifts previously inverted~\cite{CMPRC43} are also
shown in these figures, corresponding to both the
effective-range~\cite{schwndt} and the
R-matrix-parametrized~\cite{stammb} fits to experimental
cross-sections. A measure of the level of uncertainty associated with
these fits may be inferred by comparing the two sets of empirical
phase shifts. The SC and CC phase shifts are mostly within a few
degrees of the empirical phase shifts, with roughly the correct
energy dependence. One sees that the differences generated by
the channel coupling differ from case to case. For $S$ waves, the SC
phase shifts happen to  correspond more closely to the empirical
values, whereas the $P$-wave resonances are only reproduced well when
coupling is included. The phase shifts for the $D$ and $F$ waves are
much smaller and are also less well known empirically. The influence
of channel coupling is strongest in the the $D_{3/2}$ phase shift
obviously because of the prominent ${3\over 2}^+$ resonance just
beyond the $h+d$ threshold. The relatively poor description of the
$D_{3/2}$ phase shifts reflects the fact that the RGM calculation
puts the $h+d$ threshold, and with that, the resonance, some 2.5 MeV
too high.

The phase shifts were evaluated by solving the RGM equations at
intervals of 0.25 MeV. However, the inversion requires phase shifts
at much more closely spaced energies to enable both the $D$ and $F$
waves to contribute without the larger energy gradient of the lower
$l$ values dominating the inversion. To allow an optimum choice of
energy bites, the RGM phase shifts were parametrized by an
effective-range expansion. We follow the prescription of Schwandt
{\em et al.\/}~\cite{schwndt}, with the correction for Coulomb
barrier penetration. For all but the $P$-wave phase shifts, the
energy dependence of each partial wave could be fitted with two
adjusted parameters. Four parameters were required by the expansion
to fit each set of $P$-wave phase shifts. The resulting parameters
give phase shifts corresponding very closely to the RGM phase shifts,
i.e. the differences are only discernible on graphs of the scale of
Fig.~\ref{fig1} and Fig.~\ref{fig2} for the $F$ waves for
$E_{\rm lab}<10$ MeV, and that is the estimated numerical accuracy of
the RGM calculations.

\subsection{Inversion potentials}
\label{subsec:potentials}

We have applied inversion to both the SC and the CC phase shifts for
$\alpha+p$ as well as $\alpha+n$ scattering. Below we present results
mainly for $\alpha+p$ scattering since the $\alpha+n$ results were
very similar.

In separate subsections we describe two ways of inverting the phase
shifts to define potentials:
\begin{enumerate}
\item Establish potentials corresponding to particular energies.
\item Establish potentials fitting particular partial waves, or
small sets of partial waves, over the entire sub-threshold energy
range.
\end{enumerate}

The first class of potentials are of immediate interest to
phenomenology and enable one to compare the energy dependence of the
RGM-derived potentials with, on the one hand, phenomenological
systematics, and, on the other, the empirical inversion
potentials~\cite{CMPRC43}. Remember that the nucleon-nucleon force is
independent of energy, and so is $v$ of
Eqs.~(\ref{hasymmb},\ref{vpot}), so that any energy dependence emerges
as a result of the non-locality of the microscopic model (including
channel-coupling effects in the CC case). The second class of
potentials provide useful comparisons with the earlier work, and also
provide, through the examination of charge symmetry properties, a
useful test of the methods.

\subsubsection{Potentials for laboratory energies of 12 MeV and 18 MeV}

Potentials have been established using mixed-case inversion for
laboratory energies of 12 and 18 MeV. More specifically, these
energies were the centres of narrow energy bites, $12\pm 0.03$ MeV
and $18\pm 0.03$ MeV.

Potentials reproducing all phase shifts at these energies behave at
large $r$ in an unphysical manner, extending much further than would
be expected and exhibiting a change in sign. See Fig.~\ref{fig4} for
the case where full channel coupling was included. This general
property previously occurred in the ``all-$l$'' fits to the empirical
phase shifts~\cite{CMPRC43}. The energy dependence deduced from
Fig.~\ref{fig4} appears reasonable over much of the radial range; as
discussed below, the volume integrals suggest a contrary behaviour.

By inversion involving the same narrow energy bites, we determined
parity-dependent potentials at 12 and 18 MeV. That is to say, we find
pairs of potentials which reproduce the values and energy dependence
of, respectively, the odd and even partial-wave phase shifts at these
energies. The parity-dependent potentials behaved reasonably in
the surface region and we present
the result for the CC case in Fig.~\ref{fig5}.
 The odd and even potentials have very different radial
forms, the odd-parity term having a much longer range, and crossing
over the even parity term, which is deeper at $r=0$. In general
terms, this is precisely the behaviour of the potential found to
reproduce the empirical phase shifts~\cite{CMPRC43}. To show this, we
express the central potential in the form $V_1(r) +(-1)^l V_2(r)$ and
compare $V_1$ and $V_2$ in Fig.~\ref{fig6} for the 12 MeV CC and SC
RGM potentials with the same quantities for empirical potentials. We
see that the channel coupling brings closer agreement, so that the CC
$V_1$ agrees very well indeed with the empirical $V_1$. The
qualitative nature of the RGM $V_2$ is the same as for the empirical
$V_2$, although the magnitude is greater. Although  Fig.~\ref{fig6}
shows only one empirical potential, it should be noted that the
R-matrix based and effective-range empirical potentials agree very
closely.

Referring back to Fig.~\ref{fig5}, we see that for $r<3$ fm both the
odd-parity and even-parity  potentials fall with increasing energy in
much the way that would be expected from systematics. The spin-orbit
potential peaks at about 1 fm, a universal property of all the
inverted potentials. However, it must be said that the spin-orbit
term is much less well determined than the central term, especially
for the $D$ waves.

As required, the even-parity potentials support a single $l=0$ bound
state. For the potential fitting the 12 MeV phase shifts, this is at
$E=-8.38$ MeV  (SC case) or at $E=-11.675$ MeV (CC case). The
odd-parity potentials support none. This $l=0$ state corresponds to
the single Pauli-forbidden state $\varphi^{(0)}$ (cf. the discussion
after Eq.~(\ref{Schrforphi})), and its appearance shows that our
potential tallies with those calculated directly from the RGM
kernels~\cite{Friedrich,Horiuchi}. We can thus be assured that the
potentials we exhibit belong to the same family as ``the true"
potentials, and not to families of their supersymmetric
partners~\cite{Baye,baye,liu}. This confirms that they can, in fact,
be regarded as approximants to ``the true" potentials.

Fig.~\ref{fig7} compares the full CC 18 MeV RGM potentials with those
for the SC RGM at the same energy. It can be seen that the coupling
to the $h+d$ channels leads to a substantial attractive effect on the
even-parity potential over the whole radial range. The effect on the
even-parity potential is probably magnified by the ${3\over 2}^+$
resonance, although, as will be seen later, this behaviour is less
discernible for the volume integrals. This is in agreement with the
12 MeV potentials, as can be deduced from Fig.~\ref{fig6}.

The energy dependence of the potentials is conveniently quantified in
terms of the volume integrals and rms radii which also facilitate
comparison with phenomenology. Table 1 presents these characteristics
for the various cases just described.

The ``all-$l$'' volume integrals and, especially, rms radii must be
treated with caution because of the large-$r$ behaviour of these
potentials. With this proviso, we note that the energy dependence of
the ``all-$l$'' volume integral is of the wrong sign just as was the
case for the empirical inverted potentials~\cite{CMPRC43}.

Table 1 also suggests that the effect of channel coupling on the
odd-parity solutions is almost as large as for the even-parity
potentials; this is not what might appear to be the case in
Fig.~\ref{fig7}. This is also a ``large-$r$'' effect, since closer
examination of Fig.~\ref{fig7} reveals that indeed coupling does have
a notable effect on the tail ($r>3$ fm) of the odd-parity potential.
It must be said that it is the ``all-$l$'' volume integrals which
approach the values expected for heavy nuclei. The parity dependence
of the  nucleon--nucleus optical potential must be small for heavy
nuclei and it might not be surprising that volume integrals
appropriate for such nuclei are roughly the mean of the odd and even
values for $^4$He.

Finally, Table 1 suggests that when the entire radial range is
considered, it is only when coupling is included that the even-parity
potential falls with energy in the expected fashion.

It is satisfying that the RGM plus inversion has led to such details
as the correct energy dependence, the spin-orbit potential peaking at
a smaller radius than the halfway point of the central potential (a
universal phenomenological property) and the expected $l$-dependence
of the channel coupling effect. Unquestionably, however, the striking
result is that the RGM phase shifts yield a potential with parity
dependence of just the character found by inverting the empirical
phase shifts.

\subsubsection{Potentials fitting all energies}
Potentials can be derived which fit the phase shifts for any
$l$-value, i.e. for each pair, $(l,j=l+\frac12)$ and
$(l,j=l-\frac12)$ for the energy range 0--20 MeV. The restricted
energy range presents certain difficulties in achieving unique
potentials, but the IP method is implemented in such a way that
{\em a priori\/} information, in particular a requirement for
minimal surface oscillations, can be incorporated. Any remaining
uncertainty can be assessed from inversions using alternative choices
of basis. In this way we find potentials that are particularly well
determined by the $l=1$ resonances. Fitting the energy-dependent
phase shifts for other $lj$ yields potentials which are less well
defined. Nevertheless, the differences between SC and CC potentials
are consistent across different choices of inversion basis.

The CC and the SC $l=1$ $\alpha$--$p$ potentials are compared in
Fig.~\ref{fig8}. The CC potential is more attractive than the SC
potential for $r>1.8$ fm and less attractive for small radii. Exactly
the same behaviour is found for the $\alpha+n$ case. Indeed, in
Fig.~\ref{fig9} we compare the $l=1$ proton and neutron SC potentials
and find very small differences. Differences between the
corresponding CC potentials are scarcely visible on a plot of this
scale. At first sight, this looks just a natural consequence of
charge symmetry, which is actually built into the RGM calculations
(in contrast to the empirical case~\cite{CMPRC43}, where charge
symmetry emerged from the experimental data). Since, however, the
Coulomb force is strong enough to put the $P$-wave resonances at
different energies for protons and neutrons, the emergence of
precisely charge-symmetric potentials confirms the correctness of the
overall RGM plus inversion calculations as well.

There is another echo of the behaviour we see in Fig.~\ref{fig8}. In
Fig.~\ref{fig7} the odd-parity potential also shows a crossover in
the sign of the coupling effect, with attraction at larger radii.
However, the effect seen in Fig.~\ref{fig7} is somewhat smaller, and
the crossover radius is less.

One can also find a potential fitting the $D$-wave phase shifts for
all energies. It is rather like the even-parity potentials shown
above, except that it is much less deep at the centre. Evidently
$S$-wave phase shifts are required to establish the behaviour near
$r=0$.

It is not obvious that one can find a single potential fitting two
$l$-values for all energies over this energy range, but a potential
can be found which simultaneously fits the $l=0$ and $l=1$ partial
waves. No other combination of $l$-values could be so fitted. In
particular, any parity-dependent potential fitting all energies is
bound to have some degree of surface  oscillations. Therefore,
representation by smooth parity-dependent potentials requires a
separate potential for each energy. This is the basis of our earlier
comment that a local potential fitting phase shifts which embody
{\em either\/} $l$-dependence {\em or\/} energy dependence
will be oscillatory in some degree.

\subsection{Perey factors and trivially equivalent potentials}
\label{subsec:Perey}
There is interest in applying local $\alpha$--nucleon potentials in,
for example, three-body models of $^6$Li~\cite{coon} and of
$^6$He~\cite{Zhukov}. In order to judge the success of such models,
it is necessary to know how important the non-locality is. Some
measure of this can be found by comparing the RGM wave functions
with those for the phase-shift-equivalent local, but possibly
$l$-dependent, potentials.

For the SC calculation we can derive a generalized Perey factor, as
introduced in  Ref.~\cite{CMNP511}. The non-local wave functions
suitable for comparison are the $\chi(\mbox{${\bf r}$})$ defined in
Eqs.~(\ref{relwf},\ref{relSchrwithpot}). The wave functions of the
equivalent local-potential problem, $\chi^{\rm local}(\mbox{${\bf
r}$})$, are derived by solving the Schr\"odinger equation with the
local potentials obtained from the inversion calculation. The
generalized Perey factor is then defined as:
\begin{equation}
R(\mbox{${\bf r}$}) = \frac{|\chi(\mbox{${\bf r}$})|}
{|\chi^{\rm local}(\mbox{${\bf r}$})|}.
\end{equation}

We present the ratio $R(\mbox{${\bf r}$})$ on the scattering plane in
Fig.\ \ref{fig10} for 18 MeV protons, with $\chi$ of the SC problem
and $\chi^{\rm local}$ of the parity-dependent potential. In this
figure the particle flux is to be imagined coming from the left, with
the impact parameter indicated along the vertical axis. The pattern
is up-down asymmetric as a result of the spin-orbit force; the ratio
is presented for the nucleon spin projection normal outwards from the
plane of the paper.

The function $R$ varies widely, from $<0.7$ to $>1.2$. Whilst there
is a strong Perey-like effect ($R< 1$) close to the centre, there are
also conspicuous regions where the non-local wave function has
considerably the greater magnitude (``anti-Perey" effect). At lower
energies the pattern is similar but slightly greater in range.
$R({\bf r})$ calculated with the wave function for the
18 MeV $l$-independent potential (see Fig.~\ref{fig4}) in the
denominator has a much wider variation, with both $R<1$ and $R>1$
regions, extending to greater radii, although the ``Perey" region
almost completely covers the region of nuclear overlap.

Since there are few partial waves, one can compare wave functions for
each $lj$ individually. In Fig.~\ref{fig11} the functions
$|\chi_{lj}(r)|$ are compared with $|\chi^{\rm parity}_{lj}(r)|$ and
$|\chi^{l{-\rm indep}}_{lj}(r)|$ calculated from the parity-dependent
and $l$-independent potentials respectively, for some of the partial
waves. As expected from the remarks above, $\chi^{l{\rm -indep}}$
differs much more from $\chi$ than $\chi^{\rm parity}$. The largest
difference between $|\chi_{lj}(r)|$ and $|\chi^{\rm parity}_{lj}(r)|$
is seen for the $S$ wave for $r<2$ fm, corresponding to the
Perey-like effect  at the nuclear centre in Fig.~\ref{fig10}.
Projecting the Pauli-forbidden state out of $\chi^{\rm parity}$ for
$l=0$ brings it somewhat closer to the RGM wavefunction for
$r < 1.5$ fm, but somewhat increases the difference beyond that
radius. One sees from Fig.~\ref{fig11} that there is a general,
but by no means universal, tendency for the RGM wave function to be
damped relative to the wave function of the preferred
(parity-dependent) local potential.

It is obvious that what we see in Fig.~\ref{fig10} is very different
from the conventional Perey damping. Both a parity-dependent local
potential and an $l$-independent non-local potential will give rise
to strong damping (or anti-damping) effects on a wave function when
compared to the phase-shift-equivalent $l$-independent local
potential. We shall now argue that the wave functions $\chi$ do
embody strong non-local effects, which are very different from what
would result from pure parity dependence. The argument hinges upon
the properties of the imaginary $\psi$-potential defined in
Eq.~(\ref{impsipot}). In Fig.~\ref{fig12} we present this quantity on
the scattering plane for the 18 MeV RGM SC wave functions. The
quantity is not calculated beyond $r=8$ fm, the matching radius. We
first note simply that this quantity differs considerably from zero,
although the asymptotic flux is conserved exactly. A non-local
potential has the property of removing flux from certain regions and
emitting it again in other regions --- hence the balance of
absorptive and emissive regions in this figure. This does not,
however, suffice to demonstrate the effect of non-locality on each
partial wave component of the wave function. The argument has to be
more subtle since even a purely real, local, parity-dependent
potential gives a $\psi$-potential with strong emissive and
absorptive regions. For example, we show in Fig.~\ref{fig13} the
imaginary $\psi$-potential constructed with the wave function of the
18 MeV SC parity-dependent local potential. It is very different from
Fig.~\ref{fig12}. The approximate antisymmetry about $z=0$ is
meaningful, as will now be discussed.

There is a striking property of ${\bf \nabla \cdot j}$: it is not
hard to show, starting from Eq.\ (10) of Ref.~\cite{MIC} and assuming
no spin-orbit potential, that a parity-dependent potential, $V_1(r)
+(-1)^lV_2(r)$, which is {\em local for each partial wave\/,} will
lead to ${\bf \nabla \cdot j(r)}$ having the form
\begin{equation}
{\bf \nabla \cdot j(r)} = \alpha \sum_{{\rm
odd}\,\, l} Z_l(r) Y_{l0}(\theta),
\label{eq:asymm}
\end{equation}
which has precise odd symmetry about the  $z=0$ plane. We have
verified this numerically as well. A consequence is that the
imaginary  $\psi$-potential should exhibit a similar property but now
modulated by the variation of $|\chi|$ over the reaction plane. The
pattern of positive and negative regions will be unaffected by
variations in $|\chi|$ and must show exact antisymmetry, although the
actual magnitude of the imaginary $\psi$-potential will not. We have
confirmed this for the imaginary $\psi$-potential calculated from the
parity-dependent inversion potentials using a judicious choice of
contour lines: the regions for which it is, respectively, $> 10^{-5}$
and $< -10^{-5}$ are precise reflections of each other in the $z=0$
plane, but the detailed ``hills and valleys" are only approximate
reflections. In fact, this is more than might have been expected,
since the derivation of Eq.~(\ref{eq:asymm}) assumes that there is no
spin-orbit force, and, indeed, ${\bf \nabla \cdot j}$ is far from
antisymmetric about $z=0$ for the case of a parity-dependent inverse
potential including spin-orbit coupling. However, the spin-orbit
force also twists the pattern of $|\chi|$ (for a nucleon with spin
quantized normal to the scattering plane), so that it differs between
the regions of positive and negative impact parameter;
for the parity-dependent inversion potential the quantity
evaluated from Eq.~(\ref{impsipot}) does have the exact antisymmetry
in sign, and approximate antisymmetry in detail.

In order to check the program, we have verified that the wave
functions from the $l$-independent potential lead to a zero imaginary
$\psi$-potential and a real spherical $\psi$-potential agreeing with
Fig.~\ref{fig4}. We have not included any real $\psi$-potentials in
this paper for reasons of space; they are highly non-spherical for
the RGM and parity-dependent cases.

Why do RGM $\chi$ wave functions not yield imaginary
$\psi$-potentials exhibiting clear antisymmetry about  $z=0$? In
Fig.~\ref{fig12}, approximate antisymmetry can be seen for $6<r<8$ fm
but not for $r<6$ fm. The explanation must lie with the fact that the
RGM kernels are not only effectively parity dependent, but also
highly non-local, partial wave by partial wave. (We stress this
because $l$-dependence is sometimes called an effective non-locality,
but this can be misleading.) Thus, inspection of the imaginary
$\psi$-potential immediately reveals the non-locality partial wave by
partial wave in addition to the effective parity dependence. In
conjunction with this, it is worth mentioning that we found the
partial-wave TELP's erratically angular-momentum dependent.

We conclude that one must use the local potentials in applications
with an awareness that non-local effects could be strong, and that
these are {\em not\/} allowed for by using parity-dependent
potentials.

It should be noted that all non-locality showing up in these
comparisons between the wave functions of the RGM and of its local
equivalents are due to the exchange effects inherent in the RGM. The
exchange effects arise from antisymmetrization as well as from the
exchange operators in the nucleon-nucleon force. The additional
non-locality caused by channel coupling can only be studied in the
behaviour of the relative wave function if the second channel is
orthogonalized to the elastic channel~\cite{SchmidSpitz}. Since our
RGM is not orthogonalized, we have only used the SC wave function
here. It is also worth mentioning that the effects of Pauli exchanges
are expected to be larger for systems of more nucleons or of more
fragments, as is shown by the well-known tendency of the eigenvalues
of the the norm operators.

\section{Discussion and conclusions}
\label{discussion}

In many ways, RGM provides the most complete framework for
calculating reactions between nuclei. Antisymmetrization, which plays
a crucial role, can be included exactly alongside channel-coupling
effects. Nevertheless, RGM is just an approximation to the few-body
(or many-body) problem, in which basis truncation has to be
accompanied by replacing the realistic nucleon-nucleon potentials
with effective ones. One can capitalize on this necessity by choosing
interactions of simple functional forms, but one should keep in mind
that there is no strict theoretical foundation of these effective
interactions. Because of this, the superiority of RGM over simpler
approaches, such as the folding model, is not unqualified, and
calculations of the folding model type
are undoubtedly more feasible
for more complex nuclei. A local folding model with
effective interactions based on local-density G-matrix theory and a
very crude representation of antisymmetrization is the most useful
current theory for heavier nuclei, although it is known to miss
certain aspects that RGM encompasses.

The emergence of parity dependence is a case in point. It has long
been known that RGM calculations predict parity dependence. We can
link this parity dependence with that of the eigenvalues $\nu_N$ of
the norm operator $\cal N$, that is, ultimately, with the Pauli
principle. The parity dependence of $\nu_N$ is most drastic and,
indeed, most trivial for a pair of identical clusters: all
$\nu_{2n+l}$ with odd $l$ are zero, and the corresponding phase
shifts are undefined, while for all even $l$ there are just finite
numbers of vanishing $\nu_{2n+l}$, and, correspondingly, all phase
shifts are well defined. In our model for the system $\alpha$+nucleon
the eigenvalues are $\nu_N=1-(-1/4)^N$, where $N=2n+l=0,$ 1,... is
the number of oscillator quanta carried by the
eigenfunction~\cite{Zaikin}. Considering that in a low-energy process
the wave function $\chi_l$ in the interaction region in each partial
wave is dominated by the component proportional to the eigenfunction
belonging to the lowest non-zero eigenvalue, the model embodied in
Eqs.~(\ref{relwf},\ref{relHam}) does carry an inherent parity
dependence. If (\ref{relSchr}) is to be replaced by an equivalent
local-potential Schr\"odinger equation, then this parity dependence
should appear in the equivalent local potential.

But how does one pin down parity dependence empirically? The quality
of fit to the scattering data is such that a wide variety of
phenomenological potentials, with and without parity dependence,
could give fits of comparable mediocre quality.  Indeed, conventional
phenomenology has failed to give a decisive answer on the existence
of parity dependence for very light systems, although the argument
for heavier cases such as $^{4}$He on $^{20}$Ne~\cite{michel} is more
convincing. For $\alpha+p$ scattering, there is insufficient
information in the data for a single energy to establish parity
dependence decisively, particularly since there has been little
reason to adopt a parametrization other than a $1 + c (-1)^l$
multiplicative factor.

By applying inversion to RGM phase shifts, we find the same  general
form of parity dependence as had been extracted from phase shifts
fitting the data over a wide range of energies (see Fig.~\ref{fig6}).
We have therefore established for the first time that RGM predicts a
form of parity dependence which, unusual as it is, is required by
data. This, it must be emphasized, has been firmly established
independently of the detailed quality of the fit to the data of the
RGM calculations. The same figure also shows that the present RGM CC
calculation leads to a potential whose parity independent component
is remarkably close to that required by data.
The fact that we can draw such
conclusions confirms that it is a useful form of phenomenology
to use inversion to reduce elastic scattering data, via phase
shifts, to local potentials.

The present work suggests that it is better to para\-metrize parity
dependence by using the general form $V_1 + (-)^l V_2$ than applying
an overall $1 + c (-1)^l$ multiplicative factor. Indeed, what we
found~\cite{CMNP517} for $^{12}$C+$\alpha$ scattering, is very
similar to the pattern of Fig.~\ref{fig6}. This suggests that a
longer ranged odd-parity component is a general property.

Local potentials have many applications beyond para\-metrizing
scattering data. If we are to attribute any meaning to the wave
functions they produce, it is important to explore their relationship
to the RGM wave functions. For this reason we have presented ``Perey"
factors, and compared our inversion potentials with trivially
equivalent local potentials. The discussion on non-locality in the
last section brings into focus the question of what local potentials
are useful for, and where they are dangerous.

It was found that the local potentials produce wave functions which
agree in broad features but differ in details appreciably from the
RGM relative wave function $\chi$. The difference is certainly
reduced if the macroscopic model is changed so that its equation of
motion may include Pauli projection \cite{Saito}. But the moderate
agreement of the wave functions in the $P$ waves shows that
refinements beyond the exclusion of forbidden states may be required.
An approach devised just for such a purpose is Schmid's fish-bone
model~\cite{fishbone}, which has been worked out both for two-body
and for three-body systems. This is still an essentially macroscopic
local-potential framework, but it takes into account the fact that
the Pauli principle not only excludes or allows certain components of
the relative wave function but also suppresses or favours them with
respect to one other.

Local potentials will nevertheless continue to be useful as {\em
phenomenological tools} in reproducing scattering data; this work has
shown how they form a link between experiment and microscopic
calculations. The non-locality, which limits their applicability, is
independent of the parity dependence, although this happens to be
also related to exchange. In short, we have shown that using a
parity-dependent potential is far from exhausting the effects of
antisymmetrization, and this could well be important for
applications.

\acknowledgments
This work was supported by the OTKA grant No. 3010 (Hungary) and by
SERC grant GR/G01522 (UK).

\figure{Twice the phase shift in radians for $S$ waves
and $P$ waves for $\alpha+p$ scattering, comparing RGM and empirical
solutions. The solid and dashed lines are the SC and CC RGM phase
shifts, respectively, and the dotted and dot-dashed lines are the
effective-range and R-matrix fits, respectively.\label{fig1}}

\figure{Twice the $D$-wave and $F$-wave phase shifts, otherwise as
for Fig.\ \ref{fig1}.
\label{fig2}}

\figure{The central and spin-orbit $\alpha+p$ potentials fitting all
partial waves, the ``all-$l$ solutions'', obtained by inverting
12 MeV (solid) and 18 MeV (dashed) RGM CC phase shifts.
Note the expanded scale in the surface region. \label{fig4}}

\figure{The central and spin-orbit $\alpha+p$ potentials fitting odd
and even partial waves separately, obtained by inverting  12
MeV and 18 MeV RGM CC phase shifts.
Solid: 12 MeV even; dashed: 18 MeV even; dots: 12 MeV odd; dot
dash: 18 MeV odd.\label{fig5}}

\figure{Comparison of the terms $V_1$, $V_2$ of the central
$\alpha+p$ potentials expressed as $V_1 +(-)^l V_2$, at 12 MeV. RGM
CC (solid line), RGM SC (dashed line), the potential inverted from
the empirical R-matrix (dotted line). \label{fig6}}

\figure{Proton-$\alpha$ central (above) and spin-orbit (below)
potentials, showing the effect of channel coupling at 18 MeV. The
solid line shows CC even, the dashed line the SC even, the dotted
line the CC odd and the dot-dashed line the SC odd potentials.
\label{fig7}}

\figure{The SC (solid line) and CC (dashed line) RGM-derived
$\alpha+p$ potentials found by inverting the $l$=1 phase shifts
over all subthreshold energies. \label{fig8}}

\figure{Comparing neutron and proton potentials:
the $l=1$ $\alpha+n$ (solid line) and $\alpha+p$ (dashed
line) ``all-energy'' potentials, SC case. \label{fig9}}

\figure{Ratio of the magnitudes of the RGM SC wave function $\chi$
and of the wave function for the parity-dependent inverted potential
for 18 MeV protons. The figure represents the scattering plane, with
the nuclear centre at the centre, the impact parameter plotted
vertically, and the protons incident from the left. The asymmetry
between the positive and negative impact parameters reflects the fact
that the nucleon spin is normal to the page. \label{fig10}}

\figure{For selected partial waves of 18 MeV protons,
magnitude of RGM radial wave function $\chi_{lj}$
(solid line) compared with magnitudes of
wave functions calculated from the parity-dependent (dashed) and
$l$-independent (dotted) phase-equivalent potentials. \label{fig11}}

\figure{The imaginary $\psi$-potential for the 18 MeV RGM SC
calculation plotted on the scattering plane as in Fig.~\ref{fig10}.
\label{fig12}}

\figure{The imaginary $\psi$-potential for the par\-ity-de\-pend\-ent
potential fitting the phase shifts of the 18 MeV RGM SC calculation,
plotted on the scattering plane as in Fig.~\ref{fig12}.
\label{fig13}}

\onecolumn
\mediumtext
\begin{table}
\caption[Table1]{Volume integrals $J_R$
and rms radii for potentials found by
mixed-case inversion at 12 MeV and 18 MeV (lab.) Characteristics of
potentials inverted from CC and SC phase shift are shown for the two
components of the parity-dependent solutions as well as for the
``all--$l$'' potentials.}
\begin{tabular}{cccccccc}
& &\multicolumn{2}{c}{Even-$l$}&\multicolumn{2}{c}{Odd-$l$}&
\multicolumn{2}{c}{All-$l$}\\
Energy & SC/CC & $J_{\rm R}$ & $<r^2>^{1/2}$& $J_{\rm R}$ &
$<r^2>^{1/2}$ & $J_{\rm R}$ & $<r^2>^{1/2}$ \\
(MeV) & & (MeV fm$^3$)& (fm)& (MeV fm$^3$)& (fm)& (MeV fm$^3$)& (fm)
\\
\hline
\multicolumn{8}{c}{Central potential}\\
12 &CC &366.9 & 1.729 & 657.7 & 2.516 & 474.9 & 2.242 \\
12 & SC&290.5 & 1.595 & 607.7 & 2.474 & - & - \\
18 & CC&347.4 & 1.744 & 620.1 & 2.460 & 479.1& 3.183 \\
18 & SC& 289.9 & 1.629 & 570.6 & 2.410 & - & - \\
\multicolumn{8}{c}{Spin-orbit potential}\\
12 & CC & 68.1 & 1.653 & 69.0 & 1.686 & 35.2 & 1.829 \\
12 & SC & 73.5 & 1.713 & 69.9 & 1.689 & - & - \\
18 & CC & 64.4 & 1.623 & 72.5 & 1.743 & 41.1 & 1.420 \\
18 & SC & 74.8 & 1.727 & 71.2 & 1.708 & -& -\\
\end{tabular}
\end{table}

\begin{references}
\bibitem[$\dagger$]{dagger} Contact address: R.MACKINTOSH@OPEN.AC.UK.
\bibitem{CMNP517} S. G. Cooper and R. S. Mackintosh, Nucl. Phys.
{\bf A517}, 285 (1990).
\bibitem{CMPRC40}S. G. Cooper and R. S. Mackintosh, Phys. Rev. C
{\bf 40}, 502 (1989).
\bibitem{CMPRC43} S. G. Cooper and R. S. Mackintosh, Phys. Rev. C
{\bf 43}, 1001 (1991).
\bibitem{CMPRC45} S. G. Cooper, M. A. McEwan, and R. S. Mackintosh,
Phys. Rev. C {\bf 45}, 770 (1992).
\bibitem{MCMNP552}M.A. McEwan, S.G. Cooper, and R.S. Mackintosh,
Nucl. Phys. {\bf A552}, 401 (1993)
\bibitem{CMInverse} S. G. Cooper and R. S. Mackintosh, Inverse
Problems {\bf 5}, 707 (1989).
\bibitem{CMNP511} S. G. Cooper and R. S. Mackintosh, Nucl. Phys.
{\bf A511}, 29 (1990).
\bibitem{CMNP513} S. G. Cooper and R. S. Mackintosh,
Nucl. Phys. {\bf A513}, 373 (1990).
\bibitem{CMZP} S. G. Cooper and R. S. Mackintosh, Z. Phys. A {\bf
337}, 357 (1990).
\bibitem{Kamimura} M. Kamimura, Prog. Theor. Phys. Suppl. {\bf 62},
236 (1977).
\bibitem{CsLK} A. Cs\'ot\'o, R. G. Lovas, and A. T. Kruppa, Phys.
Rev. Lett. {\bf 70}, 1389 (1993).
\bibitem{BLPR} G. Bl\"uge and K. Langanke, Phys. Rev. C {\bf 41},
1191 (1990).
\bibitem{BLFB} G. Bl\"uge and K. Langanke, Few-Body Systems {\bf 11},
137 (1991).
\bibitem{CTT} F. S. Chwieroth, Y. C. Tang, and D. R. Thompson, Phys.
Rev. C {\bf 9}, 56 (1974).
\bibitem{RT} I. Reichstein and Y. C. Tang, Nucl. Phys. {\bf A158},
529 (1970).
\bibitem{HH} P. Heiss and H. H. Hackenbroich, Phys. Lett. {\bf 30B},
373 (1969).
\bibitem{PLNGyV} K. F. P\'al, R. G. Lovas, M. A. Nagarajan, B.
Gyarmati, and T. Vertse, Nucl. Phys. {\bf A402}, 114 (1983).
\bibitem{LKL} R. G. Lovas, A. T. Kruppa, and J. B. J. M. Lanen,
Nucl. Phys. {\bf A516}, 325 (1990).
\bibitem{Saito} See e.g. S. Saito, Prog. Theor. Phys. Suppl. {\bf
62}, 11 (1977).
\bibitem{HoriuchiPerey} H. Horiuchi, Prog. Theor. Phys. {\bf 71}, 535
(1984).
\bibitem{Timm} W. Timm, H.-R. Fiebig, and H. Friedrich, Phys. Rev. C
{\bf 25}, 79 (1982).
\bibitem{Schmid} E. W. Schmid, A. Faessler, H. Ito, and G. Spitz,
Few-Body Systems {\bf 5}, 45 (1988).
\bibitem{Friedrich} H. Friedrich, Phys. Rep. {\bf 74}, 210 (1981).
\bibitem{Horiuchi} H. Horiuchi, Trends in Nuclear Physics,
Vol. 2, ed. P. J. Ellis and Y. C. Tang, (Addison-Wesley, New York,
1991) p. 277.
\bibitem{SchmidNP} E. W. Schmid, Nucl. Phys. {\bf A416}, 379c (1984).
\bibitem{LovasPal} R. G. Lovas and K. F. P\'al, Nucl. Phys. {\bf
A424}, 143 (1984).
\bibitem{Buck} B. Buck, H. Friedrich, and C. Wheatley, Nucl. Phys.
{\bf A275}, 246 (1977).
\bibitem{Baye} D. Baye, Phys. Rev. Lett. {\bf 58}, 2738 (1987).
\bibitem{liu}Q.K.K. Liu, contribution to International conference on
Inverse Scattering, Bad Honnef, May 1993 (to be published by
Springer-Verlag)
\bibitem{Kanada} H. Kanada, T. Kaneko, S. Nagata, and M. Nomoto,
Prog. Theor. Phys. {\bf 61}, 1327 (1979).
\bibitem{Howell} L.L. Howell, S. A. Sofianos, H. Fiedeldey, and G.
Pantis, Nucl. Phys. {\bf A556}, 29 (1993).
\bibitem{MIC} R. S. Mackintosh, A. A. Ioannides, and S. G. Cooper,
Nucl. Phys. {\bf A483}, 173 (1988).
\bibitem{imago} S. G. Cooper and R. S. Mackintosh, User's manual for
IMAGO, Preprint OUPD9201.
\bibitem{schwndt} P. Schwandt, T. B. Clegg, and W. Haeberli,  Nucl.
Phys. {\bf A163}, 432, (1971).
\bibitem{stammb} Th. Stammbach and R. L. Walter, Nucl. Phys.
{\bf A180}, 225, (1972).
\bibitem{baye}D. Baye, contribution to International conference on
Inverse Scattering, Bad Honnef, May 1993 (to be published by
Springer-Verlag)
\bibitem{coon}N.W. Schellingerhout, L.P. Kok, S. A. Coon, and R. M.
Adam, preprint 1993.
\bibitem{Zhukov} M. V. Zhukov, B. V. Danilin, A. A. Korsheninnikov,
D. V. Fedorov, L. V. Chulkov, J. S. Vaagen, and J. M. Bang, Proc.
Int. Symp. on Structure and Reactions of Unstable Nuclei, Niigata,
1991, ed. K. Ikeda and Y. Suzuki (World Scientific, Singapore, 1991)
p. 158.
\bibitem{fishbone} E. Schmid, Z. Phys. A {\bf 297}, 105 (1980); {\em
ibid.} {\bf 302}, 311 (1981).
\bibitem{SchmidSpitz} E. Schmid and G. Spitz, Z. Phys. A {\bf 321},
581 (1985).
\bibitem{Zaikin} D. A. Zaikin, Nucl. Phys. {\bf A170}, 584
(1971).
\bibitem{michel}F. Michel and G. Reidemeister, Z. Phys. A. {\bf 333},
331, (1989).
\end{references}
\end{document}